\documentclass[prb,aps,superscriptaddress]{revtex4}

\usepackage{graphicx}
\usepackage{amssymb}
\usepackage{amsthm}
\usepackage{amsmath}
\usepackage{bbold}
\usepackage{mathtools}
\usepackage{color}

\newcommand*{\I}{\mathrm{i}}

\newcommand*{\lft}{\!\left}

\newcommand{\ket}[1]{\left|#1\right\rangle}

\def\ginn{\gamma_\inn}
\def\gout{\gamma_\out}
\def\inn{{\rm in}}
\def\out{{\rm out}}
\newcommand*{\sig}[2]{\sigma^#1(#2)}
\newcommand*{\sigl}[3]{\sigma_#1^#2(#3)}
\newcommand*{\vsigl}[2]{{\vec\sigma}_#1(#2)}
\newcommand*{\chil}[2]{\chi_#1(#2)}
\newcommand*{\etal}[2]{\eta_#1(#2)}
\newcommand*\hl[2]{h_#1(#2)}
\newcommand*\hlo[1]{h(#1)}
\newcommand*\chilo[1]{\chi(#1)}
\newcommand*\etalo[1]{\eta(#1)}

\usepackage{ulem}

\begin{document}

\title{Emulating Majorana fermions and their braiding by Ising spin chains}

\author{Stefan \surname{Backens}}
\affiliation{Institut f\"ur Theorie der Kondensierten Materie, Karlsruhe Institute of Technology,
D-76131 Karlsruhe, Germany}
\author{Alexander \surname{Shnirman}}
\affiliation{Institut f\"ur Theorie der Kondensierten Materie, Karlsruhe Institute of Technology,
D-76131 Karlsruhe, Germany}
\affiliation{Institute of Nanotechnology, Karlsruhe Institute of Technology, D-76344
Eggenstein-Leopoldshafen, Germany}
\author{Yuriy \surname{Makhlin}}
\affiliation{Condensed-matter physics Laboratory, National Research University Higher
School of Economics, 101000 Moscow, Russia} 
\affiliation{Landau Institute for Theoretical Physics, acad.~Semyonov av. 1a, 142432,
Chernogolovka, Russia}
\author{Yuval \surname{Gefen}}
\affiliation{Department of Condensed Matter Physics, Weizmann Institute of Science, 76100 Rehovot,
Israel}
\author{Johan E. \surname{Mooij}}
\affiliation{Kavli Institute of Nanoscience, Delft University of Technology, 2628 CJ Delft, The
Netherlands}
\author{Gerd \surname{Sch\"on}}
\affiliation{Institut f\"ur Theoretische Festk\"orperphysik, Karlsruhe Institute of Technology,
D-76131 Karlsruhe, Germany}
\affiliation{Institute of Nanotechnology, Karlsruhe Institute of Technology, D-76344
Eggenstein-Leopoldshafen, Germany}

\date{\today}

\begin{abstract}
We analyse the control of Majorana zero-energy states by mapping the fermionic system onto a chain of
Ising spins. 
Although the topological protection is lost for the Ising system, the mapping  
provides additional insight into the nature of the quantum states.
By controlling the local magnetic field, one can separate the Ising chain 
into ferromagnetic and paramagnetic phases,
corresponding to topological and non-topological sections of the fermionic system.
In this paper we propose (topologically non-protected) protocols performing the braiding operation, 
and in fact also  more general rotations. We first consider a T-junction geometry, but we also propose a protocol for a purely one-dimensional system.
Both setups rely on an extra spin-1/2 coupler.
By including the extra spin in the T-junction geometry, we overcome limitations due to the 1D  character of the Jordan-Wigner transformation. 
In the 1D geometry the coupler, which controls one of the Ising links, 
 should be manipulated once the ferromagnetic (topological) section of the chain
is moved far away. 
We also propose experimental implementations of our scheme. One is based on a chain of flux qubits
which allows for all needed control fields. We also describe how to translate our scheme for the 1D
setup to a chain of superconducting wires  hosting each a pair of Majorana edge states.
 
\end{abstract}

\maketitle

\section{Introduction}

Theoretically Majorana zero-energy states (also known as Majorana fermions) arise as a property of
the Kitaev chain~\cite{KitaevMajorana}.
A physical realization is provided by a one-dimensional (1D) p-wave superconductor or -- more accessible to
experiment -- by  a semiconducting wire with strong spin-orbit interaction and proximity-induced 
superconductivity~\cite{OregPRL105.177002}. A fundamental elementary quantum gate, the
braiding operation~\cite{NayakReview},  can be performed in these systems if one uses a T-junction
geometry~\cite{AliceaNatPhys7}. This adiabatic manipulation (errors emerging in such manipulations are discussed in Ref.~\onlinecite{Knapp2016})
allows for a restricted class of operations only; in order to construct a universal set of gates, it
is necessary to expand the set of gates by non-adiabatic manipulations (for discussion and further 
references see Ref.~\onlinecite{Karzig2016}).

Mathematically closely related is a 1D Ising chain, which can be mapped onto the Kitaev chain 
via the Jordan-Wigner transformation. The ideal systems are exactly equivalent, however, they differ 
in a crucial way in the presence of disorder~\cite{Greiter2014}, e.g., due to random fields. The Majorana states in the
Kitaev chain 
are topologically protected (with exponential accuracy for long chains).
I.e., local perturbations due to disorder do not lift the degeneracy of the ground states. Similarly the system remains topologically protected during the adiabatic braiding operation. In
contrast,  for the Ising chain a local longitudinal magnetic field does lift the degeneracy. 

In spite of the lack of topological protection, simulating the physics of Majorana
fermions by artificially constructed Ising chains can provide additional insight (see, e.g., the proposals of Refs.~\onlinecite{TserkovnyakPhysRevA.84.032333,Mezzacapo13}). 
Short Ising chains exhibiting Majorana physics have already been simulated in optical systems~\cite{Xu2016,Xu2017}.
With present-day Josephson qubit circuits, which have reached an encouraging level of coherence, the
Majorana physics could now be studied for interesting length and time scales. Several proposals
on how to emulate Majorana states with Josephson 
circuits have already been formulated (see, e.g., Refs.~\onlinecite{LevitovMooij,vanHeckNJP14}), but as
 a consequence of the 1D  character of the Jordan-Wigner transformation the extension to a T-junction geometry, required for performing the braiding operation, remains problematic. 
 
 
In this paper we first present a  setup of Ising spins in a T-junction geometry which 
-- except for the topologically protection --
is equivalent to the fermionic Majorana system and allows performing  the braiding operation.
It relies on an extra spin-1/2 controlling the three-spin coupling in the junction.
This spin also introduces Klein factors which assure that the fermionic anticommutation relations for different legs of the setup are obeyed.
In addition, we  propose a scheme based on a purely 1D geometry, which allows performing an operation which is equivalent to the braiding. 
This is achieved by attaching an extra spin-1/2 (the coupler) to the chain, such that one of the
links of the Ising chain is controlled by this spin. Depending on the quantum state of the coupler,
this link can be either ferromagnetic or antiferromagnetic, or a superposition thereof. The coupler
 should be manipulated 
once the ferromagnetic part of the chain, corresponding to the topological part hosting the Majorana fermions, has been moved sufficiently far away. 
An added advantage of this scheme is that it allows performing parity-conserving $U(1)$ rotations
of the Majorana qubit 
by an  arbitrary angle, whereas the topologically protected braiding in the fermionic system
fixes this angle to discrete multiples of $\pi/4$. In this way one can
construct a universal set of quantum logical gates.

Furthermore, we propose experimental implementations of our 1D setup and braiding scheme.
One is based on a chain of Josephson flux qubits. They are well suited because they can be strongly coupled and  the matrix element which plays the role of the transverse field can be efficiently tuned~\cite{Paauw2009}.  
We also propose  a direct translation of the 1D Ising scheme  into the realm of topological
wires~\cite{Milestones,Hell2016} hosting pairs of Majorana edge states, which allows arbitrary rotations of the Majorana qubit.

Throughout this paper we frequently use terminology appropriate for fermionic systems with topological protection, e.g., referring to topological qubits and braiding operations. This is done to emphasize the equivalence of the fermionic system and the Ising chains. We do not imply the existence of topological protection in Ising systems. But we want to stress -- and this is one of the main messages of the present paper -- that due to the exact mapping between the fermionic and Ising systems, topological notions have exact counterparts in the qubit chain.

\section{Ising chain}
\label{sec:IsingChain}

Let us first recall some well-known relations between Ising chains and Majorana fermions. 
We consider an Ising chain described by the Hamiltonian~\cite{PFEUTY197079}
\begin{align}
\label{eq:HIsing}
H &= - \sum_{n=1}^N \hlo{n} \, \sig{x}{n} - J \, \sum_{n=1}^{N-1} \sig{z}{n} \, \sig{z}{n+1}\ .
\end{align}
For definiteness we consider the ferromagnetic case, $J>0$. 
If the perpendicular magnetic field is weak, $\hlo{n} < J$,
the ground state of a long chain,  $N\rightarrow \infty$, is ferromagnetic; for strong fields the
system is paramagnetic. 
By controlling the transverse fields $\hlo{n}$ at each site we can define a 
ferromagnetic domain, which is an interval where $\hlo{n} \ll J$, in an otherwise paramagnetic chain where $\hlo{n} \gg J$. In the paramagnetic state, at low temperature considered here, the paramagnetic spins are "frozen" in the state $\ket{\rightarrow_x}$, the
($+1$)-eigenstate of $\sigma^x$. The two ``ground states'',  $\ket{\uparrow\uparrow\uparrow}$ and
$\ket{\downarrow\downarrow\downarrow}$, of the ferromagnetic section are 
degenerate. 
If $\hlo{n}=0$ for all $n$ in the chain 
the degeneracy is exact~\footnote{Interestingly, the degeneracy between the 
somewhat perturbed ground states remains exact even if $h(n)$ vanishes only at a single site, whereas 
in the rest of the chain $\hlo{n} \ll J$.}. For non-vanishing but small $\hlo{n}$ a residual hybridisation between these two states decays exponentially as a function of the length of the ferromagnetic domain~\cite{PFEUTY197079}. This remains so also when the ferromagnetic domain is surrounded by polarized paramagnetic domains
 (for further discussions see Ref.~\onlinecite{Narozhny2017}). 
By switching the
transverse fields $\hlo{n}$ on and off one can adiabatically control the size and position of  the
ferromagnetic interval,  similar to the ``zipping" and ``unzipping" procedure described in
Ref.~\onlinecite{Dorner2003}.

 By means of the Jordan--Wigner transformation 
\begin{align}
\chilo{n} &= \sig{z}{n} \, \prod_{p=1}^{n-1} \sig{x}{p} \ ,\label{eq:chin}\\
\etalo{n} &= \sig{y}{n} \, \prod_{p=1}^{n-1} \sig{x}{p} \ ,\label{eq:etan}
\end{align}
we can map the Ising model to a fermionic system with Hamiltonian
\begin{align}
\label{eq:HMajorana}
H &=- \I \, \sum_{n=1}^N \hlo{n} \, \chilo{n} \, \etalo{n} - 
	 \I \, J \, \sum_{n=1}^{N-1} \etalo{n} \, \chilo{n+1}
\end{align}
and  fermionic Majorana operators $\chilo{n}$, $\etalo{n}$ satisfying the anticommutation relations
\begin{align}
\{ \chilo{n}, \chilo{m} \}_+ &= 2 \, \delta_{nm} 
\nonumber\\
\{ \etalo{n}, \etalo{m} \}_+ &= 2 \, \delta_{nm} \
\nonumber\\
\{ \chilo{n}, \etalo{m} \}_+ &=0\ .
\end{align}
Furthermore, by introducing the local Dirac fermions $a(n) \equiv [\etalo{n} + \I \chilo{n}]/2$ and
$a^\dag(n) \equiv [\etalo{n} - \I \chilo{n}]/2$ we recover the
Kitaev-Hamiltonian~\cite{KitaevMajorana} of a 1D p-wave superconductor 
\begin{equation}
H =  \sum_{n=1}^N  \hlo{n} \, \left[2 a^\dag(n) a(n) - 1\right] - J \, \sum_{n=1}^{N-1} 
\left[ a^\dag(n) a(n+1) + a(n)a(n+1) + {\rm h.c.} \right]\ .
\label{Kitaev}
\end{equation}
We observe that the perpendicular magnetic field of the Ising system is equivalent to the chemical
potential of the Kitaev model~\cite{KitaevMajorana},
$-2h(n) \, \hat = \, \mu(n)$, and the ferromagnetic coupling strength $J$ replaces the hopping
matrix element $w$, which in the case described by Eq.~(\ref{Kitaev}) is chosen to coincide with the
gap of the p-wave superconductor, $J \,  \hat = \,  w =\Delta$ (see
Ref.~\onlinecite{KitaevMajorana}).

In the Hamiltonian (\ref{eq:HMajorana}) we note that for $-2 h(n) \, \hat = \, \mu(n) = 0$ in the whole chain, the two boundary operators $\gamma_L\equiv \chilo{1}$ and
$\gamma_R\equiv\etalo{N}$ do not appear. They represent the famous zero-energy Majorana
modes~\cite{KitaevMajorana}. 
For non-vanishing but weak fields, $0 \ne h(n) \ll J$, 
the Majorana modes are no longer perfectly localized but acquire a finite extent. However, the overlap vanishes exponentially with growing length of the system. This is the origin of the topological protection of the Majorana system.
For the Majorana system it is usual to introduce  the 
Dirac fermion $d_0 = (\gamma_L - \I \gamma_R)/2$ and the parity operator of the zero energy subspace
$P_0\equiv -\I  \gamma_R \gamma_L= 1- 2 d_0^\dag d_0^{\phantom \dag}$.  Its even and odd eigenstates, 
$P_0\ket{0}=\ket{0}$, $P_0\ket{1}=-\ket{1}$,
satisfy
$d_0^\dag \ket{0} = \ket{1}$, $d_0 \ket{1} = \ket{0}$, $d_0^\dag \ket{1} = d_0 \ket{0} =0$.

In the spin system for vanishing perpendicular field the two states 
$\ket{\uparrow\uparrow\uparrow}$ and $\ket{\downarrow\downarrow\downarrow}$
are degenerate and span a qubit space.
 For non-vanishing but weak fields the spin states are no longer perfectly ferromagnetic. However, if the ferromagnetic domain is long the ensuing hybridisation is exponentially suppressed.
The mapping between the Majorana and the spin system is achieved by $\gamma_L=\chilo{1}
= \sig{z}{1}$ and $\gamma_R=\etalo{N}=\sig{y}{N} \prod_{p=1}^{N-1} \sig{x}{p} 
=-\I \sig{z}{N} P$, where the total parity operator of the chain reads $P\equiv \prod_{p=1}^{N} \sig{x}{p}$. 
Thus we have 
\begin{align}
&\gamma_L \ket{\uparrow\uparrow\uparrow} = \ket{\uparrow\uparrow\uparrow}
\nonumber\\
&\gamma_L  \ket{\downarrow\downarrow\downarrow} = -\ket{\downarrow\downarrow\downarrow}\
\nonumber\\
&\gamma_R \ket{\uparrow\uparrow\uparrow} =  \I \ket{\downarrow\downarrow\downarrow}
\nonumber \\
&\gamma_R \ket{\downarrow\downarrow\downarrow} =- \I \ket{\uparrow\uparrow\uparrow}\ .
\end{align}
I.e., $\gamma_L \, \hat = \, \tau^z$ and $\gamma_R \, \hat = \,  \tau^y$ correspond to two non-commuting 
spin-like operators (Pauli matrices)  in the 2-dimensional subspace of ferromagnetic states.
From here we also find the mapping between the even and odd eigenstates of the Majorana system and the ferromagnetic states of the spin system
\begin{align}
&\ket{0} \equiv \frac{\ket{\uparrow\uparrow\uparrow} + \ket{\downarrow\downarrow\downarrow}}{\sqrt{2}} \nonumber\\
&\ket{1} \equiv \frac{\ket{\uparrow\uparrow\uparrow} - \ket{\downarrow\downarrow\downarrow}}{\sqrt{2}}
\end{align}
Finally we mention that the non-topological state of the Majorana system, found for $|\mu| \ge
2|w|$, corresponds to the paramagnetic phase of the spin system, $|h|\ge|J|$.

{\color{blue} }

\section{T-Junction with 3 Ising chains}

\begin{figure*}
\begin{center}
\includegraphics[width=0.4\textwidth]{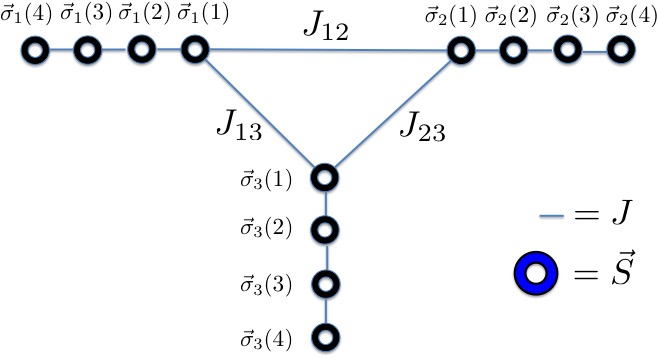}\quad\quad\quad
\includegraphics[width=0.4\textwidth]{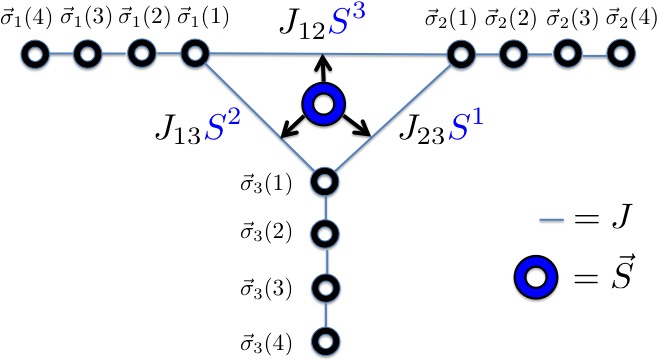}
\end{center}
\caption{\label{fig:3chains} Three Ising chains. {\bf Left panel}: Ising coupling between the chains. Here, a fictitious 
spin $\vec S$ is introduced formally in order to construct the Klein factors. (See Text.)
{\bf Right panel}: A central spin $\vec S$ controls 
the coupling between the chains via a 3-spin interaction.}
\end{figure*}

For strictly 1D setups the mapping between an Ising chain and the Majorana system is exact. 
However, for the latter the braiding operation has been recognized as an essential tool,
which can be achieved in a T-junction geometry. In order to emulate this operation
we consider now 3 Ising chains coupled to each other as shown in Fig.~\ref{fig:3chains}. 
The Hamiltonian of each of the chains with label $\alpha = 1,2,3$ and -- for convenience -- the site index $n$
counting from the junction, reads
\begin{align}
H_\alpha &=- \sum_{n=1}^N \hl{\alpha}{n} \, \sigl{\alpha}{x}{n} - J \, \sum_{n=1}^{N-1}
\sigl{\alpha}{z}{n} \, \sigl{\alpha}{z}{n+1}\ .
\end{align}

One could try to extend the Jordan--Wigner transformation by introducing 
a global 1D ordering, i.e., numbering the sites in all three chains in some order. However, any such order is artificial
and leads to non-local interactions,  thus precluding mapping to a fermionic system with a
quadratic Hamiltonian and braiding by established procedures~\cite{NoriBraiding}.
An alternative is provided by a modified Jordan-Wigner transformation similar to the one 
proposed in Ref.~\onlinecite{Crampe2013526}. It introduces an extra spin, denoted by the Pauli matrices $S^\alpha$,
\begin{align}
\label{eq:JWSchi}
\chil{\alpha}{n} &=S^\alpha\, \sigl{\alpha}{z}{n} \, \prod_{p=1}^{n-1} \sigl{\alpha}{x}{p}\ , \\
\label{eq:JWSeta}
\etal{\alpha}{n} &= S^\alpha\,\sigl{\alpha}{y}{n} \, \prod_{p=1}^{n-1} \sigl{\alpha}{x}{p}\ .
\end{align}
This extra spin provides Klein factors. 
The Hamiltonian of each chain is again of the form (\ref{eq:HIsing}), while due to the extra 
operators $S^\alpha$ also the Majorana operators belonging to different chains anticommute. Thus, we have 
properly ``fermionized" the whole system. In what follows it will be important to know the commutation 
relations between $S^\alpha$ and the Majorana operators (\ref{eq:JWSchi}) and (\ref{eq:JWSeta}),
\begin{align}
\label{eq:SMajoranaCommutation}
[S^\alpha,\chil{\alpha}{n}]_-=0 \quad &, \quad [S^\alpha,\etal{\alpha}{n}]_-=0\ , \nonumber \\
\{S^\alpha,\chil{\beta}{n}\}_+=0 \quad &, \quad \{S^\alpha,\etal{\beta}{n}\}_+=0\ \quad{\rm for\
}\alpha\ne\beta.
\end{align}
This doubles the Hilbert space and 
creates effectively two equivalent copies of the system. With this extension one arrives at a simple fermionic description. 

To proceed one could introduce additional Ising couplings between the first spins of each chain, as
indicated in the left panel of Fig.~\ref{fig:3chains}, with Hamiltonian 
\begin{align}
\label{eq:HintIsing}
H_{\rm int}^{\rm Ising} &= - \sum_{\alpha < \beta} J_{\alpha\beta}\,
\sigl{\alpha}{z}{1} \sigl{\beta}{z}{1}\ .
\end{align}
In the fermionic representation this becomes
\begin{eqnarray}
\label{eq:HintIsingF}
H_{\rm int}^{\rm Ising} 
= \I  \sum_{\alpha < \beta}
J_{\alpha\beta}\, \chil{\alpha}{1}\, \chil{\beta}{1}\,
\epsilon^{\alpha\beta\gamma}\,S^\gamma \ .
\end{eqnarray}
This form  suggests a connection to the Kondo model, which was 
explored in Refs.~\onlinecite{PhysRevLett.110.147202,Pino2015} and is most interesting for critical Ising chains 
with $h \approx J$. This connection may not be obvious, since some spin and
fermionic operators do not commute with each other, cf.~Eq.~(\ref{eq:SMajoranaCommutation}). To
overcome this obstacle, one can proceed with the modified Jordan-Wigner transformation~\cite{Crampe2013526}.
Since this approach is  irrelevant for our purposes it will not be discussed here any further.
Furthermore, we note that the simple Ising coupling (\ref{eq:HintIsing})  between the chains does not
lead to a quadratic Hamiltonian in terms of Majorana fermions.
This will influence protocols intended to simulate braiding-type protocols in
such systems. 
In fact, as can be seen in the spin description, the braiding-type procedure 
used in Ref.~\onlinecite{NoriBraiding} does not produce a desired
qubit operation, rather it may lead to an identity operation.

As an alternative we suggest proceeding with another coupling Hamiltonian, as illustrated in  the right panel of 
Fig.~\ref{fig:3chains}.
It relies on a three-spin interaction with an extra spin-1/2 with components $S^\alpha$ in the junction.
We call this extra spin the ``coupler". The interaction between the chains is described by the
Hamiltonian 
\begin{equation}
\label{eq:HintS}
H_{\rm int}^{S} = - \frac{1}{2} {\sum_{\alpha\beta\gamma}}'
J_{\alpha\beta} S^\gamma \sigl{\alpha}{z}{1}\, \sigl{\beta}{z}{1}
\end{equation}
with summation over mutually distinct $\alpha$, $\beta$, $\gamma$ and
$J_{\alpha\beta}=J_{\beta\alpha}$.
The fermionic version of this coupling reads
\begin{equation}
\label{eq:HintSFermi}
H_{\rm int}^{S} = \frac{\I}{2} \sum_{\alpha\beta\gamma} \epsilon^{\alpha\beta\gamma}
J_{\alpha\beta}\, \chil{\alpha}{1}\, \chil{\beta}{1} \,, 
\end{equation}
with the Levi-Civita symbol $\epsilon^{\alpha\beta\gamma}$.
Thus the coupling~(\ref{eq:HintS}) leads to a quadratic fermionic Hamiltonian; cf.~a
similar discussion in Ref.~\onlinecite{Crampe2013526} and three-spin couplings employed in
Refs.~\onlinecite{Xu2016,Xu2017}.
This ensures that braiding can be performed by standard methods~\cite{AliceaNatPhys7}. 
Note that the operators $S^\alpha$ disappear from the Hamiltonian. Nevertheless, 
as we will see below, the spin is not decoupled since the components $S^\alpha$ do not, in general, commute with
$H_{\rm int}^{S}$ (cf.~Eqs.~(\ref{eq:SMajoranaCommutation})).

To simulate braiding we follow the braiding protocol for a T-junction geometry as described in Ref.~\onlinecite{AliceaNatPhys7}. 
We begin with the state with only one  ferromagnetic  domain  (representing the topological section) placed in the left leg, $\alpha=1$.
In the first stage of the operation the ferromagnetic  domain is transported
adiabatically through the upper link of the coupler region to the right leg, $\alpha=2$, then through the
right link to the lower leg, $\alpha=3$, and finally through the left link back to the left leg.
At each stage of the protocol only one of three couplings $J_{\alpha\beta}$ is switched on while the ferromagnetic 
($\sim$topological) domain  is  transported through the coupler.  Let
us consider the effect of these operations, first in the spin language. During the first stage, if
$S^3=+1$, the state of
the ferromagnetic region (qubit) is conserved, while for $S^3=-1$ all spins are switched between up- and down. 
To proceed it is useful to move to the Heisenberg picture and record what happens to the operators.
We define in the (low-energy) qubit space the Pauli matrix operators $\hat\tau^{x,y,z}$,
in the up-down ($\ket{\uparrow\uparrow\uparrow}$, 
$\ket{\downarrow\downarrow\downarrow}$) basis. In the
Heisenberg picture, during the first stage $\tau^x$ and $S^3$ are conserved, while
$\tau^{y,z}\to S^3\tau^{y,z}$ and $S^{1,2}\to \tau^xS^{1,2}$. In the second and third stages
transformations are similar, with the part of $S^3$ played by $S^1$ and $S^2$, respectively. 
The combination of all three stages yields
\begin{align}
\tau^z \rightarrow  S^3 \tau^z \rightarrow \left(\tau^x S^1 \right) S^3 \tau^z  
\rightarrow S^2 \left(\tau^x S^1 \right) S^3 \tau^z  &=& - \tau^y
\nonumber\\
\tau^y \rightarrow  S^3 \tau^y \rightarrow \left(\tau^x S^1 \right) S^3 \tau^y  
\rightarrow S^2 \left(\tau^x S^1 \right) S^3 \tau^y  &=&  \tau^z \, 
\end{align}
whereas $\tau^x$ and the three coupler spin components $S^1$, $S^2$, $S^3$ are preserved. 
After the three stages the coupler spin is again disentangled from the qubit, although they were entangled during the process,
and the braiding operation is achieved (i.e., a $\pi/2$ $x$-rotation
in the language of pseudospin $\boldsymbol\tau$). Indeed, within the doubly degenerate ground state the two fermionic Majorana
operators, $\ginn=\tau^z$ and $\gout=\tau^y$ are interchanged (see
Eqs.~(\ref{eq:chin},\ref{eq:etan}) and after Eq.~(\ref{eq:HIsing})): $\ginn\to\gout$,
$\gout\to-\ginn$. (Since the notation $\gamma_L$ and $\gamma_R$, used in Section~\ref{sec:IsingChain} for Majorana zero-modes becomes ambiguous in the T-junction geometry we use here
the notation $\ginn$ for the Majorana zero-mode which is closest to the junction and $\gout$ for the one that is further away.)

We conclude that the coupling~(\ref{eq:HintS}) allows simulating the braiding protocol of Majorana systems in a spin system, which is one of the main messages of this paper. 
On the other hand, we can not exclude that the coupling is difficult to realize in an experiment. 
Fortunately, there are further possibilities, including one which is based on a strictly 1D geometry, 
which will be discussed in the following section.

\section{A braiding-type protocol in a 1-D chain geometry}

\begin{figure*}
\begin{center}
\includegraphics[width=0.5\textwidth]{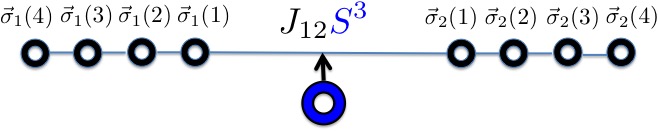}
\end{center}
\caption{\label{fig:2chainsSpin} Two Ising chains coupled via the central spin $\vec S$.}
\end{figure*}

Based on the considerations presented above we suggest an alternative approach replacing the braiding, based on
a purely one-dimensional geometry as displayed  in Fig.~\ref{fig:2chainsSpin}. 
I.e., we consider a system with only two chains ($\alpha=1$
and $2$), with a 3-spin interaction coupling the two chains, 
$H_{\rm int}^{S} = - J_{12} \sigl{1}{z}{1} \sigl{2}{z}{1}\,S^{3}$.
Effectively, the transport of the ferromagnetic ($\sim$topological) domain through all three components
$S^1$, $S^2$, and $S^3$,
described for the T-junction geometry, can be replaced by transport through a single spin provided we 
rotate the state of the coupler between various stages of the protocol. 
Depending on the angles of these rotations, the following protocol also provides a generalization of the single-domain braiding protocol, namely allowing for an arbitrary rotation angle, cf.~below. Moreover, a single intermediate rotation of the coupler spin is
sufficient, as we will demonstrate.

The braiding protocol in terms of spins described in the previous section showed the following:
 When the ferromagnetic 
 interval is adiabatically transported 
through the coupler, the spins of the domain remain intact if the coupling is ferromagnetic
($S^\gamma=+1$) and flip if the coupling is antiferromagnetic ($S^\gamma=-1$). If the coupler is in
a superposition of the two states, the coupler and the ferromagnetic domain become entangled.
In the 1D geometry, one again should first pull the ferromagnetic region (qubit) through
the special Ising link which is controlled by the coupler spin  $S^3$. The stages controlled by
$S^1$- and $S^2$ are no longer available, which limits the accessible possibilities. However, one
could emulate various coupler-spin components by rotations of the coupler spin between the pulling
operations. Thus, for instance, one can achieve  a braiding-type operation 
by pulling the qubit through the coupler
three times with proper coupler-spin rotations in between. The rotations should be chosen in such a
manner that three spin components 
are involved in sequence. This statement can be also confirmed by
a direct calculation of the evolution of $\boldsymbol\tau$ and $\bf S$ operators. Moreover, one can
achieve a a braiding-type operation also  with a shorter manipulation, which we will describe next.

Indeed, the full braiding, which we  just outlined, is achieved after a sequence
of three pulling operations past the coupler spin components  $S^3$, $S^1$, and $S^2$. However, we notice that
the operator $S^2$ returns to its initial value after the first two stages (this follows from the
observation that all three coupler-spin components return back after the full cycle, and $S^2$ is
conserved during the last, $S^2$-pulling stage). This allows us to avoid the third ($S^2$-)pulling by
using the following approach: if the coupler is initially prepared in the ($+1$)-eigenstate of $S^2$,
then $S^2=+1$ during the third pulling operation, and thus this operation is trivial (the qubit
spins keep their states) and can be left out. Hence, the third stage is not needed, simplifying the
overall procedure. 

In the Appendix A we describe  our suggested protocol in detail; here we concentrate of the underlying ideas
and main result. 
One Ising link is controlled by the $S^3$ component of the coupler-spin. In
order to emulate other components, we manipulate the coupler spin via local fields. To minimally
disturb the qubit state, this should be done while the ferromagnetic ($\sim$topological) region is moved away from the coupler
location. Let us follow the fate of the qubit and coupler operators in the three stages of the process: 
the first and third stage involve pulling the ferromagnetic section past the coupler, while the second stage involves a
coupler-spin rotation $S^2\to S^2$, $S^1\to S^1\cos\theta+S^3\sin\theta$, $S^3\to
S^3\cos\theta-S^1\sin\theta$. After the combination of the three stages we find that $S^2$ is
conserved, while $\tau^y\to\tau^y\cos\theta-\tau^z S^2\sin\theta$ and $\tau^z\to\tau^z\cos\theta
+\tau^y S^2\sin\theta$. In other words, if the coupler was initially prepared, e.g., in the (+1)
eigenstate of $S^2$ (cf.~the next section), it remains in this state after the
operation, that is, the coupler remains disentangled from the qubit. At the same time a `braiding
rotation' is performed on the qubit subspace: $\ginn\to\ginn\cos\theta + \gout\sin\theta$,
$\gout\to\gout\cos\theta - \ginn\sin\theta$.

In Appendix \ref{sec:BraidingMajoranaLanguage} we reformulate the braiding-type protocol 
 in  Majorana fermionic representations.

\section{Physical realisation of the 3-spin coupling}

An effective coupling of the type $H_{\rm int}^{S} = - J_{12} \sigl{1}{z}{1}
\sigl{2}{z}{1}\,S^1$ 
(which is equivalent to that considered above upon the substitution $S^1 \leftrightarrow S^3$) 
can be realised as follows. Consider three spins $\boldsymbol\sigma_1$, $\boldsymbol\sigma_2$, and
$\bf S$ coupled by the following Hamiltonian 
\begin{align}\label{eq:3SpinH}
H =  v_1 \sigma_{1}^z S^z + v_2 \sigma_{2}^z S^z - \Delta S^x\ ,
\end{align} 
where $\Delta \gg v_1,v_2$. Under these conditions one can treat the couplings $v_1$ and $v_2$
perturbatively 
with $H_0 = - \Delta S^x$ and  perturbation  
$V=(v_1 \sigma_{1}^z + v_2 \sigma_{2}^z) S^z$. 

A Schrieffer-Wolf transformation $\tilde H = e^{-R}\,H\,e^{R}$ with
\begin{align}\label{eq:SWtrafo}
R = -\frac{\I S^y}{2\Delta}(v_1 \sigma_{1}^z + v_2 \sigma_{2}^z)\ .
\end{align}
yields $[H_0, R]_- = -V$. We thus find the new Hamiltonian 
\begin{align}
\tilde H \approx H_0 +\frac{1}{2}\,[V,R]_-=- \Delta S^x
  - \frac{S^x}{2\Delta}(v_1 \sigma_{1}^z + v_2 \sigma_{2}^z)^2=
-\tilde \Delta S^x - \frac{v_1 v_2}{\Delta}\, \sigma_{1}^z\, \sigma_{2}^z \,S^x  \ ,
\end{align} 
where $\tilde \Delta = \Delta + \frac{v_1^2+v_2^2}{2\Delta}$.
This produces the desired interaction with $J_{12} = \frac{v_1 v_2}{\Delta}$ if we transform to the rotating frame with respect to the term 
$- \tilde \Delta S^x$.

In a realistic situation we have 
to include the terms corresponding to the transverse fields acting on the spins $\boldsymbol\sigma_1$,
$\boldsymbol\sigma_2$, i.e., $-h_1 \sigma_1^x$ and $-h_2 \sigma_2^x$ into $H_0$. A straightforward
calculation shows that our conclusions remain intact as long as $\Delta \gg h_{1,2}$.  
Formally, also the limit $\Delta \ll h_{1,2}$ is possible (the effective coupling is then of the
order $\Delta v_1 v_2/h_{1,2}^2$). However, in this limit, once the spins
$\boldsymbol\sigma_1$, $\boldsymbol\sigma_2$ are being ``frozen" and ``defrozen", one would have to go
through the resonances $\Delta = h_{1,2}$, where unwanted transitions would happen. Thus, we require the condition
$\Delta \gg h_{1,2}$ at all stages of the protocol. One can also show that the terms describing the
coupling of $\boldsymbol\sigma_1$ and $\boldsymbol\sigma_2$ to the further spins of the respective chains as well
as the driving term $-\Omega(t)S^z$ acting on the coupler are only slightly modified by the
transformation (\ref{eq:SWtrafo}).      

\section{Practical implementation of the Ising chain}

\begin{figure*}
\begin{center}
\includegraphics[width=0.8\textwidth]{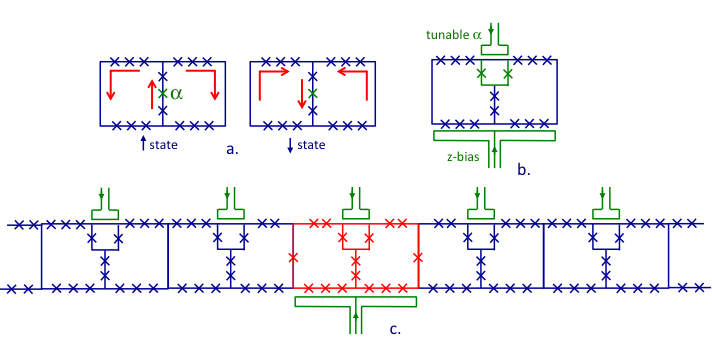}
\end{center}
\caption{\label{fig:Implementation}Practical realisation of qubits chains with Josephson junction
circuits. {\bf a}: Basic states of the gradiometer flux qubit. A permanent flux of half a flux
quantum is applied to each loop. Tunnelling occurs through one smaller junction denoted by $\alpha$. {\bf b}: Driving of
the qubit by changing of the tunnelling strength ($\sigma^x$) or the flux bias ($\sigma^z$). {\bf
c}: Layout of the system with two chains (left and right) and the coupler (center).}
\end{figure*}
A practical implementation of the proposed schemes can be realised by means of Josephson junction circuits. Flux qubits
are particularly suited because a strong coupling between them is easy to achieve and  their
tunnelling strength, which plays here the role of the transverse field, can be efficiently
tuned~\cite{Paauw2009}. 
A possible circuit is illustrated in Fig.~\ref{fig:Implementation}. It consists of a chain of
gradiometer flux qubits~\cite{Paauw2009}. An individual
gradiometer qubit is shown in Fig.~\ref{fig:Implementation}a. It is an advanced version of the flux
qubit~\cite{Mooij1036}
(persistent current qubit with a characteristic smaller junction denoted by $\alpha$), in which the
outer inductance loop is replaced by two symmetrically placed loops. The two states denoted by
$\ket{\uparrow}$ and $\ket{\downarrow}$ (eigenstates of $\sigma^z$) are the persistent current states
depicted in Fig.~\ref{fig:Implementation}a. The amplitude of tunnelling between $\ket{\uparrow}$ and
$\ket{\downarrow}$,
denoted here by $h$, is controlled by the Josephson energy of the $\alpha$-junction, which can be
tuned by replacing it with a two-junction SQUID loop (see Fig.~\ref{fig:Implementation}b). The
gradiometer geometry allows tuning the $\alpha$-junction without affecting the bias energy between
$\ket{\uparrow}$ and $\ket{\downarrow}$. Typical values could be: qubit tunnelling strength $h=2\pi
\hbar \times 1$GHz within the topological region and $h=2\pi \hbar \times 9$GHz outside. The
gradiometer inductive loops contain further Josephson junctions; the strength of the
nearest-neighbour coupling can be varied by their number. The nearest-neighbour coupling could be
$J=2\pi \hbar \times 3$GHz. 

The coupler is a special gradiometer qubit with a particularly strong tunnelling, denoted here by
$\Delta$ (see Eq.~(\ref{eq:3SpinH})). We envision $\Delta=2\pi \hbar \times 15$GHz. Moreover, an
extra magnetic bias ($z$-bias) is needed in order to perform NMR-like rotations of the coupler. This
bias can be described 
by adding to the Hamiltonian a term $H_z = -\Omega(t) S^z$. The coupling between the coupler and 
the first qubits of the chains can be made as strong as $v_{1/2}=2\pi \hbar \times 8$GHz (see
Eq.~(\ref{eq:3SpinH})).
This is achieved by placing extra Josephson junctions into the legs mutual
to the coupler and the first qubits (Fig.~\ref{fig:Implementation}c).
These numbers are given as examples, they can be realised with available technology.

Switching of the individual qubits of the chains as well as of the coupler can be effected in a few 
nanoseconds while the coherence time can be above $10{\rm\mu s}$~\cite{Yan2016}. This will allow at least 1000 operations, enough to slide the topological region (5 qubits long) through the coupler and back. 

Simulating a controllable Ising-Kitaev chain in the way described above would validate the high level of coherence and control of superconducting flux qubits. This would demonstrate that the superconducting qubits are sufficiently advanced and may simulate fermionic dynamics.

\section{Implementation in a topological system with Majorana edge states.}

\begin{figure*}
\begin{center}
\includegraphics[width=0.4\textwidth]{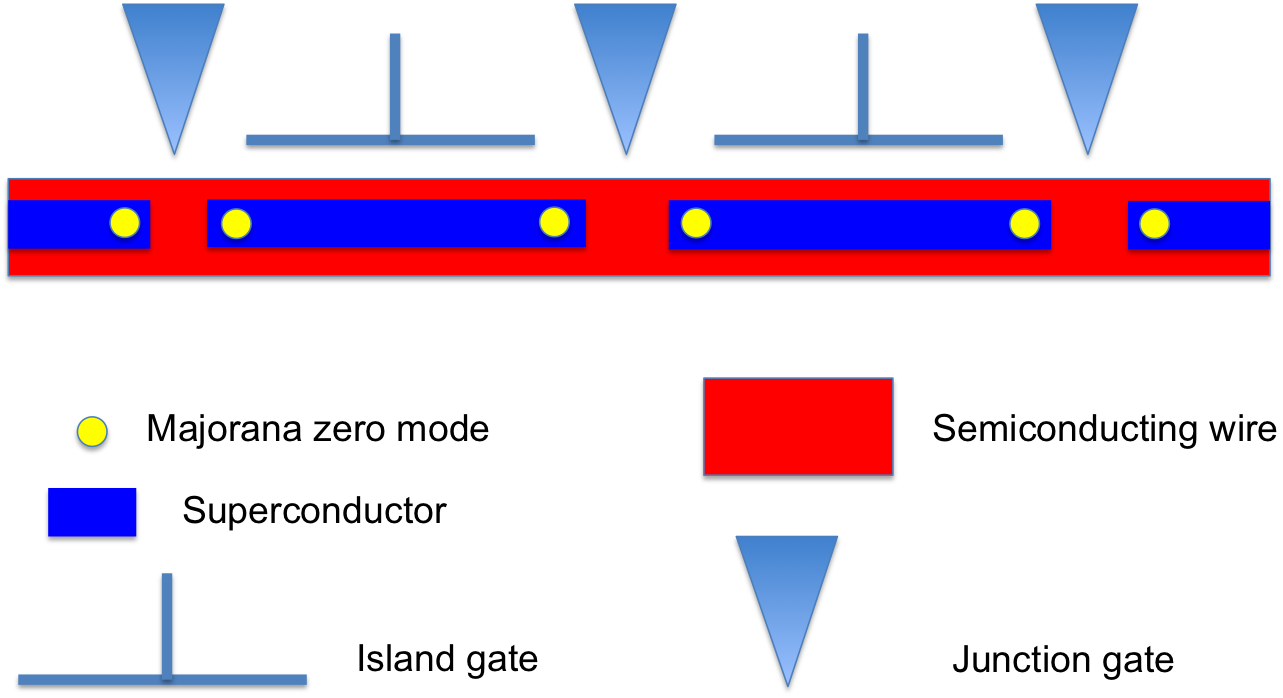}
\end{center}
\caption{\label{fig:MajoranaChain1}Implementation of a Kitaev chain using superconducting islands 
on top of a semiconducting wire (see Ref.~\onlinecite{Milestones}). The gates control both the electro-chemical potential of each wire (transverse field in the Ising equivalent) as well as the tunnelling amplitude between the adjacent Majoranas (analog of $J$ in the Ising chain). }
\end{figure*}

We now ask the following question: Do the protocols developed in this paper for the Ising spin systems provide added insight into the use of topological systems with fermionic Majorana zero modes? First, we mention the recent developments (see Refs.~\onlinecite{Milestones,Hell2016}) with advanced designs of Majorana wires. Instead of small semiconducting wires placed on top of large superconductors one uses  multiple, relatively small elongated superconducting islands placed on top of a semiconducting wire. In Fig.~\ref{fig:MajoranaChain1} we show a chain of such islands (as described in detail  in Refs.~\onlinecite{Milestones,Hell2016}). It is easy to understand that by allowing for tunnelling between adjacent Majoranas fermions, 
$H_T= i \sum_n E_M  \etalo{n} \, \chilo{n+1}$, we obtain a system completely equivalent to the Kitaev chain. Here $\etalo{n}$ is the right Majorana fermion of island $n$,  $\chilo{n}$ is the left Majorana fermion of island $n$, and $E_M$ is the tunnelling amplitude discussed in detail in Ref.~\onlinecite{Hell2016}. The charging effects are suppressed if the 
residual Josephson coupling between the islands, $E_J$, which is always present in addition to Majorana coupling $E_M$ (see Refs.~\onlinecite{Milestones,Hell2016}), is larger than the charging energy, $E_C$ of each island. Alternatively, one can think of coupling the islands to large superconducting reservoirs~\cite{Milestones}. The capacitive interactions between the islands do not change the situation 
qualitatively, as long as they remain small~\cite{Gangadharaiah2011}.
\begin{figure*}
\begin{center}
\includegraphics[width=0.4\textwidth]{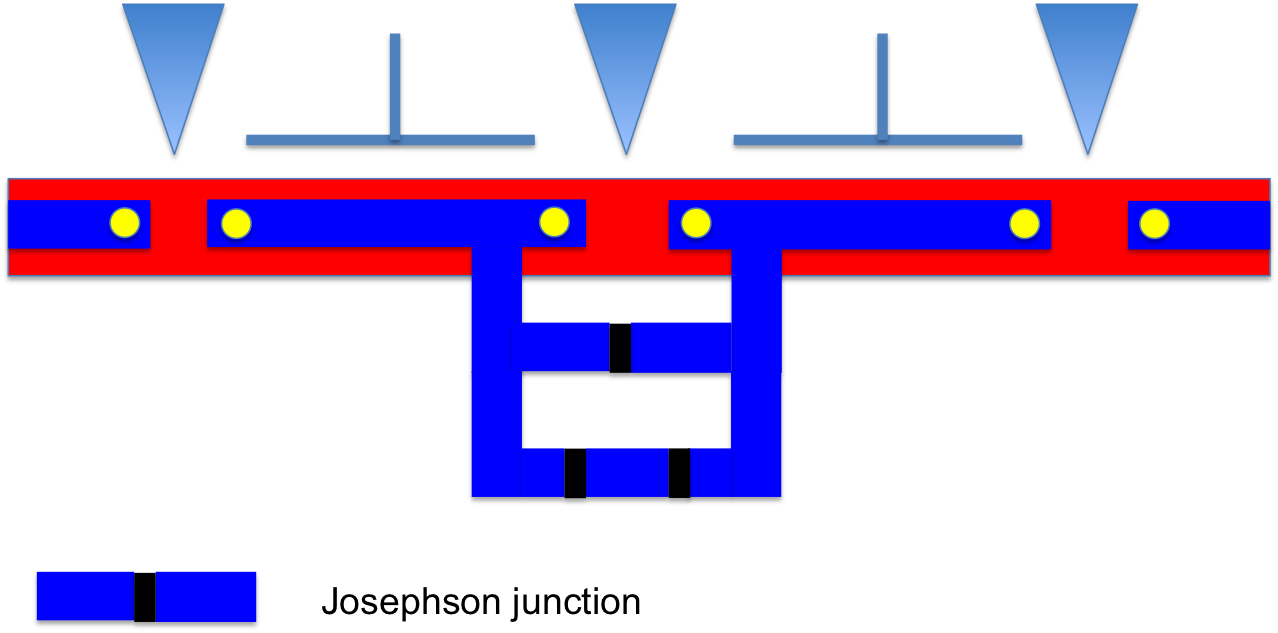}
\end{center}
\caption{\label{fig:MajoranaChain2}A flux qubit controls one of the links in the Kitaev chain. Assuming the Josephson energy of the qubit dominates over the tunnelling amplitude of Majoranas, the latter is enslaved to the quantum state of the qubit.}
\end{figure*}

With the insight gained from the spin representation we envision coupling a flux qubit
to one of the links between adjacent sections hosting Majorana fermions as shown 
in Fig.~\ref{fig:MajoranaChain2}. 
At this stage we do not discuss the experimental realisation of this setup, 
which may present a challenge for present-day technology.
The idea is to use the phase difference $\varphi$ across the Josephson junction of the qubit (the
branch with a single junction) to control the tunnelling of Majoranas. 
As discussed, e.g., in Ref.~\onlinecite{Hell2016}, the tunnelling Hamiltonian for this particular link reads $H_T = i E_M \cos(\varphi/2) \etalo{n} \, \chilo{n+1}$. We consider the situation where the Josephson energy of the flux qubit is much higher than the tunnelling amplitude $E_M$. Then the phase difference across the junction can take two well defined values, depending on the quantum state of the flux qubit, and the system of Majoranas has to adjust to these two possible situations. In a usual symmetrically biased flux qubit the phase drop $\varphi$ takes the values $\varphi = \pi \pm \Delta\varphi$ with $\Delta\varphi < \pi/2$. 
Thus, $\cos(\varphi/2)$ takes one positive and one negative value, and the effective Hamiltonian of the link controlled by the flux qubit becomes $H_{\rm eff} = i J' S^z \etalo{n} \, \chilo{n+1}$, where $J' \equiv E_M |\cos(\pi/2 + \Delta\varphi/2)|$. This Hamiltonian is equivalent to the one considered above for the 1D setup.

\section{Discussion}

In this paper we have proposed setups of Ising spins which allow emulating the physics of Majorana systems 
and, in particular, performing the braiding operation. 
In the first case we showed that this is possible in a T-junction geometry, when we introduce an extra 
spin coupling to the three legs and also appearing as Klein factor in the generalized Jordan-Wigner transformation, thus
preserving the proper anticommutation relations.
In the second case we showed that a braiding-type protocol can also be performed  in a strictly 
1D Ising setup, which may be easier to realize in experiments. In addition it allows performing
the analogue of parity conserving rotations 
of a Majorana qubit by an arbitrary angle. This is achieved by placing a special spin-1/2 (coupler)
in one of the Ising links of the chain. The coupler controls the sign (ferro- or
anti-ferromagnetic) of that Ising link. 

This 1D setup emerged in the search for methods to
implement (non-protected) topological braiding in Ising chains. The well-known problem is the fact that the
Jordan-Wigner transformation, which maps the Ising problem onto the fermionic one, is strictly
one-dimensional. Placing a coupler between three Ising chains allows circumventing this obstacle but 
requires a very special kind of coupling. We showed that the same can be achieved in 1D if one is 
able to manipulate the coupler between the adiabatic passages of the ferromagnetic (topological) domain
through the Ising link controlled by the coupler. 

Our protocol could possibly be generalized. One prospective possibility is to rotate the coupler 
during sliding of the topological (ferromagnetic) domains. This could induce non-trivial Berry phases.
Moreover, our protocol can be directly employed in systems containing 
topologically protected fermionic wires (islands). Such wires would have to be coupled 
in a non-protected fashion to a spin-1/2 (qubit).

\section{Acknowledgements}
This work was supported by DFG and RSF under RSF grant No.~16-42-01035 and DFG grant
No.~SH~81/4-1 as well as DFG grant No.~SH~81/3-1, DFG CRC183, and
the Minerva foundation.

\begin{appendix}
\section{Braiding algorithm in the spin language}
\label{sec:BraidingSpinLanguage}

In this appendix we complement the previous discussions by an explicit calculation in the spin 
language. We assume that the chains $\alpha=1$ and $2$ are on the left and right side of the setup,
and the coupling between the chains is controlled by $S^3$. We perform the following
protocol:
(i) We start with the ferromagnetic  ($\sim$ topological) section (qubit) on the left side prepared in the state
$\ket{\uparrow\uparrow\uparrow}$, while the coupler is in the state $\ket{\uparrow}_S$. I.e.,  the initial state is the product state
\begin{equation}
\ket{\psi_0} = \ket{\uparrow}_S \otimes \ket{\uparrow\uparrow\uparrow} \ .
\end{equation}
(ii)  To rotate the coupler around the $x$-axis by an angle $\phi$,  we apply the operation 
\begin{equation}
U_\phi = \exp\left\{-\frac{i\phi }{2}S^x\right\} = \cos\frac{\phi}{2} - i \sin\frac{\phi}{2}S^x\ .
\end{equation}
This gives 
\begin{equation}
\ket{\psi_1} = U_\phi \ket{\uparrow}_S \otimes \ket{\uparrow\uparrow\uparrow} = 
\left[\cos\frac{\phi}{2} \ket{\uparrow}_S  - i
\sin\frac{\phi}{2} \ket{\downarrow}_S\right] \otimes \ket{\uparrow\uparrow\uparrow} \ .
\end{equation}
(iii) We pull the ferromagnetic section adiabatically to the right side. This gives 
\begin{equation}
\ket{\psi_2} =\cos\frac{\phi}{2} \ket{\uparrow}_S \otimes \ket{\uparrow\uparrow\uparrow} - i
\sin\frac{\phi}{2} \ket{\downarrow}_S\otimes \ket{\downarrow\downarrow\downarrow} \ .
\end{equation}
(iv) We rotate the coupler around the $y$-axis by an angle $\theta$, by applying
\begin{equation}
U_\theta = \exp\left\{-\frac{i\theta}{2} S^y\right\} = \cos\frac{\theta}{2} - i
\sin\frac{\theta}{2} S^y\ .
\end{equation}
This produces
\begin{equation}
\ket{\psi_3} =\cos\frac{\phi}{2} \left[\cos\frac{\theta}{2}\ket{\uparrow}_S +
\sin\frac{\theta}{2} \ket{\downarrow}_S\right] \otimes \ket{\uparrow\uparrow\uparrow} -
i
\sin\frac{\phi}{2} \left[\cos\frac{\theta}{2} \ket{\downarrow}_S -
\sin\frac{\theta}{2} \ket{\uparrow}_S\right]\otimes
\ket{\downarrow\downarrow\downarrow} \ .
\end{equation}
(v) We pull the ferromagnetic section back to the left side. This produces
\begin{eqnarray}
\ket{\psi_4} &=&\left[\cos\frac{\phi}{2} \cos\frac{\theta}{2} \ket{\uparrow}_S -
i \sin\frac{\phi}{2} \cos\frac{\theta}{2} \ket{\downarrow}_S\right]
\otimes \ket{\uparrow\uparrow\uparrow} \nonumber\\
&+& \left[\cos\frac{\phi}{2} \sin\frac{\theta}{2} \ket{\downarrow}_S + i
\sin\frac{\phi}{2} \sin\frac{\theta}{2} \ket{\uparrow}_S\right] \otimes
\ket{\downarrow\downarrow\downarrow}\ .
\end{eqnarray}
For $\phi = \pi/2$ (this corresponds to a preparation of an eigenstate of $S^2$ at step (ii) above)
the resulting state is a product state 
\begin{eqnarray}
\ket{\psi_4} &=&\frac{1}{\sqrt{2}}\left[\ket{\uparrow}_S - i  \ket{\downarrow}_S\right] \otimes 
\left(\cos\frac{\theta}{2} \ket{\uparrow\uparrow\uparrow} +
i\sin\frac{\theta}{2}\ket{\downarrow\downarrow\downarrow}\right)\ .
\end{eqnarray}

If, instead, we started with the state $\ket{\psi'_0} =  \ket{\uparrow}_S \otimes
\ket{\downarrow\downarrow\downarrow}$,
we would obtain 
\begin{eqnarray}
\ket{\psi'_4} &=&\frac{1}{\sqrt{2}}\left[\ket{\uparrow}_S - i  \ket{\downarrow}_S\right] \otimes 
\left(\cos\frac{\theta}{2} \ket{\downarrow\downarrow\downarrow} +
i\sin\frac{\theta}{2} \ket{\uparrow\uparrow\uparrow} \right)\ .
\end{eqnarray}
Thus we perform a $(-\theta)$-rotation around the $x$-axis of the  qubit, and after the
operation the coupler is no longer entangled with 
the qubit. In the basis of the states with well-defined occupation numbers (parity) of the Dirac
fermions, $(1/\sqrt{2})(\ket{\uparrow\uparrow\uparrow} \pm \ket{\downarrow\downarrow\downarrow})$,
the achieved rotation is a $U(1)$ rotation around the $z$-axis, as 
expected~\footnote{The frozen spins are in the state $\ket{\rightarrow_x} =
(1/\sqrt{2})(\ket{\uparrow}+\ket{\downarrow})$. If this spin "joins" 
the $\ket{\uparrow\uparrow\uparrow}$ chain the rotation is performed by 
$U=\exp\left[\I (\pi/4)\,\sigma_y\right]$. We have $U\ket{\rightarrow_x} = \ket{\uparrow}$. If the
frozen spin "joins" the $\ket{\downarrow\downarrow\downarrow}$ chain the rotation 
is by $U^\dag=\exp\left[-\I (\pi/4)\,\sigma_y \right]$. We obtain $U^\dag \ket{\rightarrow_x} = 
\ket{\downarrow}$.
Thus, no extra sign appears.}.

\section{Braiding in terms of the Majorana operators}
\label{sec:BraidingMajoranaLanguage}
\subsection{Boundary translations}

Assume that the topological ($\sim$ferromagnetic) interval is placed on one side of the coupler (that is, in one of the
chains). 
The shift of the inner boundary of the interval towards the coupler, say, 
from spin $\vsigl{\alpha}{k+1}$ to spin $\vsigl{\alpha}{k}$ due to the adiabatic
variation of the corresponding transverse field $\hl{\alpha}{k}$ from $+\infty$ to $0$ is given by
the operator 
\begin{align}
T_{k,\alpha}^{\inn}=\exp\lft\{ \I \, \tfrac{\pi}{4} \, \sigl{\alpha}{z}{k+1} \,
\sigl{\alpha}{y}{k} \right\} =
\frac{1+\I \, \sigl{\alpha}{z}{k+1} \, \sigl{\alpha}{y}{k}}{\sqrt{2}} = \frac{1
- \chil{\alpha}{k+1} \, \chil{\alpha}{k}}{\sqrt{2}} \ .
\end{align}
In particular, the zero-mode edge operator is shifted properly:
\begin{align}\label{eq:chiTransport}
T_{k,\alpha}^{\inn}\, \chil{\alpha}{k+1} \,T_{k,\alpha}^{\inn,\dag} =  \chil{\alpha}{k}\,.
\end{align}
The part of the boundary zero mode $\ginn$ is played by $\chil{\alpha}{k+1}$ before the shift
and by $\chil{\alpha}{k}$ after the shift.
Similarly, the operator that shifts the outer boundary away from the coupler, i.e., 
from spin $\vsigl{\alpha}{k}$ to spin $\vsigl{\alpha}{k+1}$, reads
\begin{align}
T_{k,\alpha}^{\out}=\exp\lft(  \I \, \tfrac{\pi}{4} \, \sigl{\alpha}{z}{k} \, \sigl{\alpha}{y}{k+1}
\right) =
\frac{1+\I \, \sigl{\alpha}{z}{k} \, \sigl{\alpha}{y}{k+1}}{\sqrt{2}} = \frac{1 - \etal{\alpha}{k}
\, \etal{\alpha}{k+1}}{\sqrt{2}} \ .
\end{align}
Again, the corresponding zero-mode operator is shifted properly:
\begin{align}\label{eq:etaTransport}
T_{k,\alpha}^{\out}\, \etal{\alpha}{k} \,T_{k,\alpha}^{\out,\dag} = \etal{\alpha}{k+1}\ .
\end{align}

If in the middle of the protocol the topological interval is shared by two chains, there are only
outer boundaries, and only the operators $T_{k,\alpha}^{\out}$ and $(T_{k,\alpha}^{\out})^{-1} =
T_{k,\alpha}^{\out,\dag}$ can be applied.
The operator that transfers the inner boundary from chain $\alpha$ to chain $\beta$ (where it
becomes an outer boundary) reads
\begin{align}
T_{\beta\leftarrow\alpha}=\exp\lft( \I \, \tfrac{\pi}{4}
\,S^\gamma \sigl{\alpha}{z}{1} \,
\sigl{\beta}{y}{1} \right) = 
\frac{1 + \Lambda_{\alpha\beta}\,\chil{\alpha}{1} \, \etal{\beta}{1}}{\sqrt{2}} \ .
\end{align}
Here $\alpha$, $\beta$, $\gamma$ are mutually distinct, and $\Lambda_{\alpha\beta}\equiv
\epsilon^{\alpha\beta\gamma}$ (which equals $\pm1$ depending on whether the shift $\alpha\to\beta$ is in
the clockwise/counter-clockwise direction).
Here we obtain an extra minus sign while shifting the zero mode, i.e., 
\begin{align}
T_{\beta\leftarrow\alpha} \chil{\alpha}{1} T_{\beta\leftarrow\alpha}^\dag = -
\Lambda_{\alpha\beta}\etal{\beta}{1}\ ,
\end{align}
which is consistent with the sign of (\ref{eq:HintSFermi}). 
As expected, the operators $S^\alpha$ do not enter final expressions in the fermionic language. Yet,
they remain important 
due to the non-trivial commutation relations (\ref{eq:SMajoranaCommutation}).

\subsection{Braiding-type operations in one-dimensional geometry}
Next we explain how to perform a braiding-type operation in the 1D geometry depicted in Fig.~\ref{fig:2chainsSpin}.
We refer to this setup as a two-chain composite.
Consider two chains $\alpha = 1$ and $\beta=2$.
We assume that, initially, the topological interval is limited by the inner boundary 
$N_\alpha^\inn \gg 1$ and the outer boundary 
$N_\alpha^{\out} \gg N_\alpha^\inn$. Thus, the zero mode is given 
by $\gamma_{\out} = \etal{\alpha}{N_\alpha^{\out}}$ and
$\ginn=\chil{\alpha}{N_\alpha^\inn}$
(due to the inverse orientation of the chain $\alpha=1$, we avoid using left and right indices).
Once the interval is transported to the chain $\beta$, the edges are located at
$N_\beta^\inn \gg 1$ and $N_\beta^{\out} \gg N_\beta^\inn$. To maintain an adiabatic regime,
we have to transfer the inner boundary in $\alpha$ to the outer boundary in $\beta$ and vice
versa. 

Now we construct the complete braiding operator:
\begin{align}
U_\text{br} &= U_R^\dagger \, U_L \, U_y(\theta) \, U_L^\dagger \, U_R\ .
\end{align}
Here the rotation of the coupler in the middle of the protocol is described by
\begin{align}
U_y(\theta) &= \exp\left(- \I \, \tfrac{\theta}{2} \, S^y \right)
	= \cos\tfrac{\theta}{2} - \I \, \sin\tfrac{\theta}{2} S^y\ . 
	\end{align}
The operator $U_R$ describes the transport of the inner edge on the left side ($\alpha=1$) to 
the outer edge on the right side ($\beta=2$):
\begin{align}
U_R = T_{\beta}^{\out}\,T_{\beta\leftarrow\alpha}\,T_{\alpha}^{\inn}\ ,
\end{align}
where 
\begin{align}\label{eq:Tinn}
T_{\alpha}^{\inn} \equiv  \prod_{k=1}^{k=N_\alpha^{\inn}-1} T^{\inn}_{k,\alpha}=
\prod_{k=1}^{k=N_\alpha^{\inn}-1} \frac{1 - \chil{\alpha}{k+1} \, \chil{\alpha}{k}}{\sqrt{2}}\ ,
\end{align}
and 
\begin{align}\label{eq:Tout}
T_{\beta}^{\out} \equiv  \prod_{k=N_\beta^{\out}-1}^{k=1} T^{\out}_{k,\beta}=
 \prod_{k=N_\beta^{\out}-1}^{k=1} \frac{1 - \etal{\beta}{k} \, \etal{\beta}{k+1}}{\sqrt{2}}\ .
\end{align}

Analogously, the operator $U_L$ describes the transport of the inner edge on the right side
($\beta=2$) to 
the outer edge on the left side ($\alpha=1$):
\begin{align}
U_L =T^{\out}_\alpha T_{\alpha\leftarrow\beta}T^{\inn}_\beta \ ,
\end{align}
where we again use the definitions (\ref{eq:Tinn}) and (\ref{eq:Tout}) with $\alpha$ and $\beta$ 
interchanged.

We obtain 
\begin{align}
U_\text{br} =  \cos\tfrac{\theta}{2}  -  \I \, \sin\tfrac{\theta}{2}
U_R^\dagger \, U_L \, S^y \, U_L^\dagger \, U_R\ .
\end{align}
Next
\begin{align}
U_R^\dagger \, U_L \, S^y \, U_L^\dagger \, U_R=
\left[T^{\out}_{\beta} T_{\beta\leftarrow\alpha} T^{\inn}_{\alpha}\right]^\dag
T^{\out}_{\alpha}\, T_{\alpha\leftarrow\beta}\, T^{\inn}_{\beta}\, S^y\,   
\left[T^{\out}_{\alpha}\, T_{\alpha\leftarrow\beta}\, T^{\inn}_{\beta}\right]^\dag\,
T^{\out}_{\beta} T_{\beta\leftarrow\alpha} T^{\inn}_{\alpha}\ .
\end{align}
Since, in our case, $S^y = S^2 = S^\beta$, we observe that $S^\beta$ commutes with 
operators $T^{\inn}_{\alpha/\beta}$ and $T^{\out}_{\alpha/\beta}$, whereas 
\begin{align}
S^\beta T_{\beta\leftarrow\alpha} = T_{\beta\leftarrow\alpha}^\dag S^\beta \quad {\rm and} \quad
S^\beta T_{\alpha\leftarrow\beta} = T_{\alpha\leftarrow\beta}^\dag S^\beta\ . 
\end{align}
This gives 
\begin{align}
U_R^\dagger \, U_L \, S^\beta \, U_L^\dagger \, U_R=&S^\beta\, 
\left[T^{\out}_{\beta} T_{\beta\leftarrow\alpha}^\dag T^{\inn}_{\alpha}\right]^\dag
T^{\out}_{\alpha}\, T_{\alpha\leftarrow\beta}^\dag\, T^{\inn}_{\beta}\,   
\left[T^{\out}_{\alpha}\, T_{\alpha\leftarrow\beta}\, T^{\inn}_{\beta}\right]^\dag\,
T^{\out}_{\beta}\, T_{\beta\leftarrow\alpha}\, T^{\inn}_{\alpha} \nonumber\\
=&S^\beta\, 
T^{\inn,\dag}_{\alpha} \, T_{\beta\leftarrow\alpha}\, T^{\out,\dag}_{\beta}\,
T^{\out}_{\alpha}\, T_{\alpha\leftarrow\beta}^\dag\, T^{\inn}_{\beta}\,   
T^{\inn,\dag}_{\beta}\,  T^{\dag}_{\alpha\leftarrow\beta}\, T^{\out,\dag}_{\alpha}\,
T^{\out}_{\beta}\, T_{\beta\leftarrow\alpha}\, T^{\inn}_{\alpha} \nonumber\\
=&S^\beta\, 
T^{\inn,\dag}_{\alpha} \, T_{\beta\leftarrow\alpha}\, T^{\out,\dag}_{\beta}\,
T^{\out}_{\alpha}\, [T_{\alpha\leftarrow\beta}^\dag]^2\,   
T^{\out,\dag}_{\alpha}\,
T^{\out}_{\beta}\, T_{\beta\leftarrow\alpha}\, T^{\inn}_{\alpha}
\ .
\end{align}
We notice that $[T_{\alpha\leftarrow\beta}^\dag]^2 =\Lambda_{\beta\alpha}\,
\etal{\alpha}{1}\,\chil{\beta}{1}$.
This immediately allows us to commute $T^{\out}_{\beta}$ and $T^{\out,\dag}_{\beta}$ out
since these consist only of $\etal{\beta}{k}$. We, thus, obtain
\begin{align}
U_R^\dagger \, U_L \, S^\beta \, U_L^\dagger \, U_R=
&S^\beta\, 
T^{\inn,\dag}_{\alpha} \, T_{\beta\leftarrow\alpha}\, 
T^{\out}_{\alpha}\Lambda_{\beta\alpha}\,\,\etal{\alpha}{1}\,\chil{\beta}{1}\,  
T^{\out,\dag}_{\alpha}\,
T_{\beta\leftarrow\alpha}\, T^{\inn}_{\alpha}\ .
\end{align}
Further, we observe that $T^{\out}_{\alpha} \etal{\alpha}{1} T^{\out,\dag}_{\alpha} =
\etal{\alpha}{N_{\alpha}^{\out}}=\gout$,
as the edge transport relation (\ref{eq:etaTransport}) suggests. We obtain
\begin{align}
U_R^\dagger \, U_L \, S^\beta \, U_L^\dagger \, U_R=
&\Lambda_{\beta\alpha}\,S^\beta\, 
T^{\inn,\dag}_{\alpha} \, T_{\beta\leftarrow\alpha}\, 
\gout\,
\chil{\beta}{1}\,   
T_{\beta\leftarrow\alpha}\, T^{\inn}_{\alpha}\nonumber\\
=&\Lambda_{\beta\alpha}\,S^\beta\, 
T^{\inn,\dag}_{\alpha} \, T^2_{\beta\leftarrow\alpha}\, T^{\inn}_{\alpha}\,
\gout\,
\chil{\beta}{1}  
\nonumber\\
=&\Lambda_{\beta\alpha}\,S^\beta\, 
T^{\inn,\dag}_{\alpha}\Lambda_{\alpha\beta}\,\chil{\alpha}{1}\,\etal{\beta}{1}\,T^{\inn}_{\alpha}\,
\gout\,
\chil{\beta}{1}  
\nonumber\\
=&-S^\beta\, 
T^{\inn,\dag}_{\alpha} \, \chil{\alpha}{1}\,T^{\inn}_{\alpha}\,
\chil{\beta}{1}\, \etal{\beta}{1}\,  
\gout
\ .
\end{align}
Finally we use (\ref{eq:chiTransport}) and obtain $T^{\inn,\dag}_{\alpha} \,
\chil{\alpha}{1}\,T^{\inn}_{\alpha}
= \chil{\alpha}{N_{\alpha}^\inn}=\ginn$. This gives 
\begin{align}
U_R^\dagger \, U_L \, S^\beta \, U_L^\dagger \, U_R
=-S^\beta\, 
\chil{\beta}{1}\, \etal{\beta}{1}\,  
\ginn\,\gout \ .
\end{align}
Since the system remains in the ground-state subspace, and after the 
completion of the protocol $\hl{\beta}{1} = +\infty$, 
we observe from (\ref{eq:HIsing}) that the first spin in the $\beta$ chain is frozen so that 
$\langle \chil{\beta}{1}\, \etal{\beta}{1} \rangle =-\I \langle \sigl{\beta}{x}{1}\rangle = -\I$. 
We, thus, finally obtain
\begin{align}
U_R^\dagger \, U_L \, S^\beta \, U_L^\dagger \, U_R
=\I\,S^\beta\, 
\ginn\,\gout
\ ,
\end{align}
and
\begin{align}
U_\text{br} =  \cos\tfrac{\theta}{2}  +
\sin\tfrac{\theta}{2} S^y \ginn\,\gout\ .
\end{align}
This result coincides exactly with that obtained using the spin representation. I.e., if the coupler 
was initially prepared in the $(-1)$-eigenstate of $S^y$, we induce a $(-\theta)$-rotation of the
topological qubit, described by 
$U_\text{br}=\cos\tfrac{\theta}{2}  -
\sin\tfrac{\theta}{2}  \ginn\,\gout =
\exp\left[-\frac{\theta}{2}\ginn\,\gout\right]$, whereas the coupler remains
disentangled from the qubit. 

\end{appendix}

\bibliographystyle{unsrt}
\bibliography{braiding}

\end{document}